\documentclass{article}
\usepackage{frascatiphys,here,graphicx,subfigure}
\begin{document}
\title{OPEN FLAVOR CHARMED MESONS}
\author{
P. C. Vinodkumar, \\
$^*${\em Department of Physics, Sardar Patel University,
  }\\{\em Vallabh Vidyanagar-388 120,Gujarat, India.} \\
Ajay Kumar Rai,         \\
{\em Department of Applied Sciences and Humanities,}\\
{\em SVNIT, Surat-395
007,Gujarat, India.}\\
$^*$Bhavin Patel, \\
 Jignesh Pandya, \\
{\em Department of Physics, Veer Narmad South Gujarat University,
  }\\{\em Surat-395 007,Gujarat, India.} \\
} \maketitle \baselineskip=11.6pt
\begin{abstract}
\leftskip1.0cm \rightskip1.0cm We present here recent results on
the investigations of the mass spectrum ( S-states and P-states),
decay constants, decay widths and life time of the D, $D_s$,and
$B_c$mesons within the framework of phenomenological potential
models.We also present the binding energy and the masses of the
di-meson molecular systems with one or more charm meson
combinations. Many of the newly found experimental open charm
states are identified with the orbital excitations of the
conventional open charm mesons while others like X(3872), Y(3930),
$D_{sJ}$(2632, 2700) \emph{etc.}, are identified as molecular like
states.

\end{abstract}
\baselineskip=14pt
\section{Introduction}
The study of  spectroscopy and the decay properties of the heavy
flavour mesonic states provides us useful information about the
dynamics of quarks and gluons at the hadronic scale. The
remarkable progress at the experimental side, with various high
energy machines such as LHC, B-factories, Tevatron, ARGUS
collaborations, CLEO etc for the study of hadrons has opened up
new challenges in the theoretical understanding of heavy flavour
hadrons. In order to understand the structure of the newly
observed zoo of open flavour meson
resonances\cite{Aubert2007,CLEO2003,PDG2006,Antimo2006} in the
energy range of 2-5 $GeV$,  it is necessary to analyze their
spectroscopic properties and  decay modes based on theoretical
models. Many of these states could be the excited charmed mesonic
states while for many other states the possibility of multi-quark
or molecular like structures are being proposed. Thus, the main
objective of the present talk includes the study of spectroscopy
and the decay properties of the open flavour charm mesons. We
study these open charm states as the excited states of the
conventional quark-antiquark systems within the frame work of a
potential model \cite{AKRai2002,AKRai2006}.

We also study, following the molecular interpretation of some of
the recently observed meson states, the binding energy and the
ground state masses of di-hadronic
molecules\cite{Colangelo2004,AKRai2007}. For the binding energy of
the di-hadronic state, we consider a large r ($r \rightarrow
\infty$) limit of the confined gluon propagator employed in our
earlier study on N-N integrations.\cite{Khadkikar1991}

\section{Theoretical methodology: A Potential Scheme}

For the light-heavy flavour bound system of $q\overline{Q}$ or
$\overline{q}Q$ we treat the heavy-quark (Q=c, b)
non-relativistically and the light-quark (q = u, d, s)
relativistically within the mesonic system. The Hamiltonian for
the case be written as
 \cite{AKRai2006}\begin{equation} \label{eq:hamiltonian}
H=M+\frac{p^2}{2m}+\sqrt{p^2+m^2}+V(r)+V_{S_{\bar {Q}} \cdot
S_q}(r) + V_{{L} \cdot S}(r)\end{equation} Where M is the heavy
quark mass, m is the light quark mass, p is the relative momentum
of each quark, V(r) is the  confined part of the quark- antiquark
potential,  $V_{S_{\bar {Q}} \cdot S_q}(r)$ and  $V_{{L} \cdot
S}(r)$ are the spin-spin and spin orbital part of the interaction.
Here we consider
 \begin{equation} V(r)=\frac{-\alpha_c}{r}+A r^{\nu}\end{equation}
where $\alpha_c=\frac {4}{3}\alpha_s$, $\alpha_s$ being the strong
running coupling constant, A and $\nu$ are the  potential
parameters. For computing the hyperfine and spin-orbit splitting,
we consider the spin dependent part of the usual OGEP  given by
 \cite{SSGershtein1995}
\begin{equation} V_{S_{\bar {Q}} \cdot S_q}(r)=\frac{2}{3}
\frac{\alpha_c}{M_{\bar{Q}} m_q} \ \vec{S_{\bar {Q}}} \cdot
\vec{S_q} \ 4\pi \delta(\vec{r}), \ \ \  V_{{L} \cdot S}(r)=
\frac{\alpha_c}{M_{\bar{Q}} m_q}\ \frac{\vec{L} \cdot
\vec{S}}{r^3} \end{equation} We employ the harmonic oscillator
wave function and use the virial theorem, to get the energy
expression from the hamiltonian defined by
Eqn.(\ref{eq:hamiltonian}). Here $\mu$ is the wave function
parameter determined using the variational method. The parameters
used here are $m_{u/d}=0.360\ GeV$,$m_{s}=0.5 \ GeV$, $m_{c}=1.41
\ GeV$, $m_{b}=4.88 \ GeV$, $\alpha_c=0.48$ (for open charm meson)
and $\alpha_c=0.36$ (for open beauty-charm meson). The computed S
and P wave mass spectrum of $D$, $D_s$ and $B_c$ mesons are
tabulated in Table  \ref{tab:1} alongwith the experimental and
other theoretical results.

\section{The decay constants and Lifetime of the open charm mesons}
The decay constant of the mesons is an important parameter in the
determination of the leptonic, non-leptonic weak decay processes.
It is related to the wave function at the origin through
Van-Royen-Weisskoff formula. Incorporating a first order QCD
correction factor, we compute them using the relation
\cite{BF1995}
 \begin{equation} \label{Eqn:4}
f^2_{P}=\frac{12\left|\Psi_{P}(0)\right|^2}{M_{P}} C^2(\alpha_s),
\
\textmd{where}C^2(\alpha_s)=1-\frac{\alpha_s}{\pi}\left[2-\frac{M_Q-m_q}{M_Q+m_q}\
ln\frac{M_Q}{m_Q} \right]\end{equation} where $M_{P}$ is the
ground state mass of the pseudoscalar states. \\

In the spectator approximation\cite{AKRai2006,alv1999} the
inclusive widths of b and c quarks decay are given by
 \begin{equation}
 \Gamma(b \rightarrow X)= \frac{ 9 \ G^2_F \left|V_{Q\bar{Q}}\right|^2
 m^5_b}{192 \pi^3}, \ \ \
 \Gamma(c \rightarrow X)= \frac{ 5 \ G^2_F \left|V_{Q\bar{q}}\right|^2
 m^5_c}{192 \pi^3}  \end{equation}
and width of the annihilation channel is computed using the
expression given by \cite{AKRai2006,alv1999}
\begin{equation}\label{ttlannrest}
  \Gamma(Anni)= \frac{G^2_F}{8 \pi}\left|V_{Q\bar{q}}\right|^2 f^2_{P} M_{P}
\sum_i m^2_i\left(1- \frac{m^2_i}{M^2_{P}}\right)^2  C_i
\end{equation} where $C_i=1$ for the $\tau \nu_{\tau}$ channel and
$C_i=3\left|V_{Q\bar{q}}\right|^2$ for $Q\bar{q}$, and $m_i$ is
the mass of the heaviest fermions. Here$\left|V_{Q\bar{q}}\right|$
and $\left|V_{Q\bar{Q}}\right|$ are the respective CKM Matrix,
where numerical values are obtained from \cite{PDG2006}. The total
width of the $Q\overline{q}$ meson decay is the addition of
partial widths \emph{i.e.} $\Gamma(total)$ $=$ $\Gamma(Q
\rightarrow X)+ \Gamma(Anni)$. In the case of the $B_c$ meson,
both the heavy quark , b and c under go the decay and the total
width is obtained as $\Gamma_{total}(B_c)=\Gamma(b\rightarrow
c)+\Gamma(c\rightarrow X)+\Gamma(Anni)$. The computed pseudoscalar
 decay constants with and without the correction factor
$C^{2}(\alpha_s)$, the total width and lifetime of $D$, $D_s$ and
$B_c$ mesons are listed
in Table \ref{tab:2} along with other model predictions and experimental values.\\
\begin{table}
\caption{\it S-Wave and P-Wave Masses (in $MeV$) }\label{tab:1}
\begin{tabular}
{|c|l|l|l|l|l|l|l|l|l|} \hline &$\nu$&$1^1S_0$&$1\ ^3S_1$&1\ $^3 P_0$&1\
$^3
P_1$&$1^1P_1$&1\ $^3 P_2$&$2^1S_0$&2\ $^3S_1$ \\
\hline \hline
D&0.5&1922&1992&2195&2203&2210&2218&2286&2294\\
&1.0&1912&1993&2347&2367&2390&2414&2580&2639\\
&1.5&1905&2003&2388&2435&2481&2527&2599&2709\\
&Expt.&1864&2006&$-$&$-$&$-$&$-$&$-$&$-$\\
&Ebert&1875&2009&2414&2438&2459&2501&2579&2629\\
&Pandya&1815&1909&2385&2417&2449&2481&2653&2690\\
$D_s$&0.5&2042&2089&2353&2364&2375&2386&2466&2476\\
&1.0&2003&2104&2512&2544&2576&2608&2813&2847\\
&1.5&1937&2135&2607&2678&2750&2821&3149&3228\\
&Expt.&1969&2112&$-$&$-$&2535&2574&$-$&$-$\\
&Ebert&1981&2111&2508&2515&2560&2569&2670&2716\\
&Pandya&2009&2110&2385&2417&2449&2481&2778&2280\\
$B_c$&1.0&6349&6373&6715&6726&6738&6749&6821&6855\\
&Lattice&6280&6321&6727&6743&6765&6783&6960&6990\\
&ALV&6356&6397&6673&$-$&$-$&6751&6888&6910\\
&EFG&6270&6332&6699&6734&6749&6762&6835&7072\\
\hline
\end{tabular}\\
Expt.\cite{PDG2006}, ALV\cite{alv1999}, EFG\cite{ebert2003},
Pandya\cite{jnp2001}, Lattice\cite{CTH1996}, Ebert\cite{ebert1998}

\end{table}

\begin{table}[t]
\caption{\it Decay constants ($f_P$) and lifetime of meson.}
\label{tab:2}
\begin{center}
\begin{tabular}{|c|c|c|c|c|l|}
\hline
System&$\nu$&$f_P$&$f_P(cor.)$&$\Gamma(total)$&$\tau$ \\
&&$MeV$&$MeV$&$10^{-4} eV$&$ps$\\
\hline \hline
$D$&0.5&231&157&6.126&1.074\\
&1.0&250&170&6.142&1.072\\
&1.5&276&187&6.167&1.067\\

&Expt.&$-$&$-$&$-$&1.040$\pm0.007$\\
&Penin&$-$&195$\pm$20&$-$&$-$\\
&Ebert&$-$&243$\pm$25&$-$&$-$\\

$D_s$& 0.5&218&156&9.148&0.719\\
&1.0&321&229&12.630&0.521\\
&1.5&451&322&18.515&0.356\\
&Expt.&$-$&$-$&$-$&0.500$\pm0.007$\\
&Heister&$-$&285&$-$&$-$\\
$B_c$& 1.0&$-$&556&13.86&0.47\\
&Expt.&$-$&$-$&$-$&0.46$^{+0.18}_{-0.16}$\\
\hline
\end{tabular}
\\Expt.\cite{PDG2006}, Ebert\cite{ebert2003}, Penin\cite{Penin2002}, Heister\cite{Heister2002}
\end{center}
\end{table}

\begin{table}[t]
\caption{\it Low-lying masses of Multiquarks as di-hadronic molecule}
\label{tab:3}
\begin{center}
\begin{tabular}{|l|l|l|l|l|l|l|l|}
\hline
Systems&J$^{PC}$&$\Omega$&$\psi$&BE&Mass&Expt\cite{PDG2006}&Others\\
$h_1-h_2$&&$GeV^2$&$GeV^{3/2}$&$GeV$&$GeV$&$GeV$&$GeV$\\
\hline \hline
$\pi$-$D$&0$^{++}$&0.0186&0.0757&0.022&2.027&$-$&$-$\\
$\pi$-$D^*$&$1^{+-}$&0.0188&0.0762&0.022&2.169&$-$&$-$\\
$K$-$D$&0$^{++}$&0.1415&0.3465&0.015&2.344&$D_{sJ}(2.317)$&$-$\\
$K$-$D^*$&$1^{+-}$&0.1455&0.3539&0.016&2.485&$D_{sJ}(2.460)$&$-$\\
$\rho$ -$D$&$1^{+-}$&0.2684&0.5602&0.033&2.603&$-$&$-$\\
$K^*$-$D$&$1^{+-}$&0.3265&0.6489&0.039&2.718&$D_{sJ}(2.700)$&$-$\\
&$0^{++}$&0.2795&0.5775&0.235&2.543&$-$&$-$\\
$\rho$ -$D^*$&$1^{++}$&$-$&$-$&0.134&2.644&$-$&$-$\\
&$2^{++}$&$-$&$-$&0.064&2.845&$-$&$-$\\
&$0^{++}$&0.3420&0.6718&0.158&2.624&$D_{sJ}(2.632)$&$-$\\
$K^*$-$D^*$&$1^{++}$&$-$&$-$&0.040&2.741&$-$&$-$\\
&$2^{++}$&$-$&$-$&0.077&2.976&$-$&$-$\\
$D$-$D$&0$^{++}$&0.3568&0.6935&0.008&3.738&$-$&3.723 \cite{Lmaiani2006}\\
$D$-$D^*$&$1^{+-}$&0.3810&0.7285&0.006&3.878&X(3.870)&3.876 \cite{Rosina2004}\\
&$0^{++}$&0.4081&0.7670&0.084&3.930&$-$&$-$\\
$D^*$-$D^*$&$1^{++}$&$-$&$-$&0.040&3.974&$-$&$-$\\
&$2^{++}$&$-$&$-$&0.048&4.062&$\psi(4.040)$&3.968 \cite{Debert2006}\\
\hline
\end{tabular}
\end{center}
\end{table}
\section{Di-hadrons as molecular states}
The low-lying di-hadronic molecular system consisting of di-meson
tetra quark states  are treated here by assuming non-relativistic.
Hamiltonian given by
\begin{equation} \label{eq:hammol} H = M + \frac{P^2}{2 \mu} +
V(R_{12})+ V_{SD}(S_1 S_2) \end{equation} where
$M=m_{h_1}+m_{h_2}$, $m_{h_1}$ and $m_{h_2}$ are masses  of the
hadrons, $\mu$ is the reduced mass, $P$ is the relative momentum
of the two hadrons and $V(R_{12})$ is the residual (molecular)
interaction potential between the two hadrons given by the
asymptotic expression ($r\rightarrow \infty$) of the confined one
gluon exchange interaction (COGEP) given by \cite{Khadkikar1991}
  \begin{equation} V(R_{12})=\frac{-k_{mol}}{R_{12}} e^{-C^2 \
R_{12}^2/2}\label{eq:confgluon1}\end{equation} where $k_{mol}$ is
the residual strength of the strong interaction coupling and $C$
is the effective colour screening parameter of the confined
gluons. Using a trial wave function given by \begin{equation}
\label{eq:wavfu} \psi(R_{12})= \left(4
\frac{\Omega^{3/2}}{\sqrt{\pi}} \right)^{1/2} e^{-\Omega \ \
R^2_{12}/2}\end{equation}  By minimizing the expectation value of
H, the ground state molecule energy is obtained as
\begin{equation}
E(\Omega)= M + \frac{3\Omega}{4 \mu} - \frac{4
k_{mol}\Omega^{3/2}}{c^2 + 2 \Omega} + \frac{8}{9} \frac{
\alpha_s}{m_{h_1} m_{h_2}}
 \ \vec S_1 \ \cdot  \vec S_2 \  |\psi(0)|^2\label{eq:varmass1}
 \end{equation}
Here, we have added the spin-hyperfine contribution separately.
The binding energy of the di-mesons as $BE=|m_{h_1}+m_{h_2}-E|$
and the parameters $k_{mol}=0.45$. and c=1.25 $GeV$ are employed
to compute the binding energy (BE) at the charmed sector. The
computed masses and binding energies of the di-meson systems are
tabulated in Table \ref{tab:3}.

\section{Conclusion and Discussion:}

The properties of open charm mesons \emph{vis a vis} $D$, $D_s$
and $B_c$ are investigated by us using an effective static
quark-antiquark interaction potential of the form
$-\frac{\alpha_c}{r}+Ar^\nu$. We found that the potential form
with $\nu=1.0$ is consistent with the experimental results of the
light-heavy flavour mesons. The relativistic treatment of light
flavour and non relativistic treatment of heavy flavour seem to be
justifiable in light of the successful prediction of the various
properties of light-heavy flavour mesons. In the case of
 $B_c$-meson study,  the non-relativistic treatment for
both the heavy quarks yields better result. The $S$-wave and
$P$-wave masses, decay constants $f_P$,the decay widths and life
time of $D$, $D_s$ and $B_c$ mesons are studied within the
potential scheme with $0.5\leq\nu < 2$. The recently observed
$D_{s1}(2536)$ and $D^{*}_{sJ}(2857)$ are found to be the $1^3P_1$
and $2^3S_1$ states predicted in our model with $\nu=1.0$ Other
predicted excited states are expected to be identified and
observed in future experiments.\\

The pseudoscalar decay constant$f_P$ predicted without the
correction terms $C^2(\alpha_s)$ of Eqn.(\ref{Eqn:4}) in our model
with potential indeax $\nu=1$ is found to be in better agreement
with the experimental values of $f_{D^+}=222.6\pm16 MeV$ of CLEO
collaboration \cite{Artuso2005}and the predicted  value of
321$MeV$ for $f_{D_s}$ is within the error bar of the experimental
result of $283\pm17\pm7\pm14$ $MeV$ by BaBar
collaboration\cite{Aubert2007}. However, the PDG average value for
$f_{D_s}$ is $267\pm33$ $MeV$ \cite{PDG2006}. The ratio of
$\frac{f_{D_s}}{f_D}$ in our case is $1.34$ with the correction
factor, while that with out correction factor is $1.28$ which is
in accordance with the Lattice results of $1.24\pm0.01\pm0.07
$\cite{Aubin2006}. The lifetime predictions of $1.07$ $ps$ for $D$
 and $0.52$ $ps$ for $D_s$ mesons are in good agreement with the respective
experimental result of $1.04\pm0.007$ ps of $D^\pm$ and
$0.5\pm0.007$ $ps$ with $\nu=1.0$.\\

 The exotic states such as
$X(3872), D_{SJ}$(2317, 2460, 2632, 2700 and $2860)$, $\psi(4040)$
etc are identified as the low lying
 di-mesonic molecular states at the charm sector as shown in Table \ref{tab:3}.
 Though there
 exist many attempts, the zoo of open flavour mesonic states continues to pose
 challenges to both experimental analysis and theoretical predictions.\\

{\bf Acknowledgement:} Part of this work is done with a financial
support from DST, Government of India, under a Major Research
Project \textbf{SR/S2/HEP-20/2006}.


\begin{thebibliography}{99}
\bibitem{Aubert2007} Aubert B et  al. (BABAR Collaboration), Phys. Rev. Lett. \textbf{98}, 122011 (2007); Phys. Rev. Lett. {\bf 97} 222001
(2006).
\bibitem{CLEO2003}D. Besson et al. (CLEO Collaboration), Phys.
Rev. {\bf D 68}, 032002 (2003).
\bibitem{PDG2006} W. M. Yao et al., (Partcle Data Group) J. Phys.
{\bf G 33}, 1 (2006).
\bibitem{Antimo2006}  Antimo Palano, Nucl. Phys. {\bf B 156},105
(2006).


\bibitem{AKRai2002} Ajay Kumar Rai, R H Parmar and P C Vinodkumar, Jnl. Phys G.
 {\bf 28},2275-2280 (2002).
\bibitem{AKRai2006}Ajay Kumar Rai and P C Vinodkumar, Pramana J.
 Phys. {\bf 66}(2006).
 \bibitem{Colangelo2004}P Colangelo et al., Mod.
 Phys. Lett.  {\bf A 19}, 2083(2004).

\bibitem{AKRai2007}Ajay Kumar Rai,  Jignesh Pandya and P C Vinodkumar, Nucl. Phys. {\bf A782}, 406-409 (2007).
\bibitem{Khadkikar1991}Khadkikar S B and Vijayakumar K B , Phys. Lett. {\bf B 254},
(1991).
\bibitem{SSGershtein1995} S. S. Gershtein et al.,Phys. Rev {\bf D51}, 3613(1995).
\bibitem{BF1995} E. Braaten and S. Fleming Phys. Rev {\bf D 52}, 181
(1995).
\bibitem{alv1999} A Abd El-Hady et al., Phys. Rev {\bf D 59}, 094001
(1999).
\bibitem{ebert2003} D. Ebert, R. N. Faustov and V. O. Galkin, Phys. Rev. {\bf D 67}, 5663
(2003).

\bibitem{jnp2001} J. N. Pandya and P. C. Vinodkumar, Pramana J. Phys. {\bf 57}, 821
(2001).

\bibitem{CTH1996} C T H Davies et al., Phys. Lett. {\bf B 382}, 131
(1996).

\bibitem{ebert1998} D. Ebert, R. N. Faustov and V. O. Galkin, Phys. Rev. {\bf D 57}, 014027 (1998).
 \bibitem{Penin2002} A. A. Penin et al., Phys. Rev. {\bf D 65}, 054006 (2002).
\bibitem{Heister2002} A. Heister et al.,ALEPH Collaboration  Phys.Lett.{\bf B
528},1(2002).
\bibitem{Lmaiani2006}   L. Maiani, et al., Phuys. Rev. {\bf D 71}, 014028 (2005).
\bibitem{Rosina2004}   D. Jane and M. Rosina, Few-Body Systems {\bf
                       35}, 175 (2004) .
\bibitem{Debert2006}   D. Ebert, R. N. Faustov and V. O. Galkin,
                       Phys. Lett. {\bf B 634}, 21 (2006).


\bibitem{Artuso2005} Artuso M et al.,Phys. Rev. Lett.{\bf 95}, 251801 (2005).
\bibitem{Aubin2006} Aubin C et al.,Phys. Rev. Lett.{\bf 95}, 122002 (2005).


\end{thebibliography}
\end{document}